\documentstyle{article}
\def\2{{1\over 2}}
\def\d{\partial}

\def\bea{\begin{eqnarray}}
\def\eea{\end{eqnarray}}

\def\beq{\begin{equation}}
\def\eeq{\end{equation}}
\def\ba{\beq\new\begin{array}{c}}
\def\ea{\end{array}\eeq}
\def\be{\ba}
\def\ee{\ea}
\def\stackreb#1#2{\mathrel{\mathop{#2}\limits_{#1}}}
\def\Tr{{\rm Tr}}

\def\f{1\over}

\makeatletter
\newdimen\normalarrayskip              
\newdimen\minarrayskip                 
\normalarrayskip\baselineskip
\minarrayskip\jot
\newif\ifold             \oldtrue            \def\new{\oldfalse}
\def\arraymode{\ifold\relax\else\displaystyle\fi} 
\def\eqnumphantom{\phantom{(\theequation)}}     
\def\@arrayskip{\ifold\baselineskip\z@\lineskip\z@
     \else
     \baselineskip\minarrayskip\lineskip2\minarrayskip\fi}
\def\@arrayclassz{\ifcase \@lastchclass \@acolampacol \or
\@ampacol \or \or \or \@addamp \or
   \@acolampacol \or \@firstampfalse \@acol \fi
\edef\@preamble{\@preamble
  \ifcase \@chnum
     \hfil$\relax\arraymode\@sharp$\hfil
     \or $\relax\arraymode\@sharp$\hfil
     \or \hfil$\relax\arraymode\@sharp$\fi}}
\def\@array[#1]#2{\setbox\@arstrutbox=\hbox{\vrule
     height\arraystretch \ht\strutbox
     depth\arraystretch \dp\strutbox
     width\z@}\@mkpream{#2}\edef\@preamble{\halign
\noexpand\@halignto
\bgroup \tabskip\z@ \@arstrut \@preamble \tabskip\z@ \cr}%
\let\@startpbox\@@startpbox \let\@endpbox\@@endpbox
  \if #1t\vtop \else \if#1b\vbox \else \vcenter \fi\fi
  \bgroup \let\par\relax
  \let\@sharp##\let\protect\relax
  \@arrayskip\@preamble}
\def\eqnarray{\stepcounter{equation}%
              \let\@currentlabel=\theequation
              \global\@eqnswtrue
              \global\@eqcnt\z@
              \tabskip\@centering
              \let\\=\@eqncr
              $$%
 \halign to \displaywidth\bgroup
    \eqnumphantom\@eqnsel\hskip\@centering
    $\displaystyle \tabskip\z@ {##}$%
    &\global\@eqcnt\@ne \hskip 2\arraycolsep
         $\displaystyle\arraymode{##}$\hfil
    &\global\@eqcnt\tw@ \hskip 2\arraycolsep
         $\displaystyle\tabskip\z@{##}$\hfil
         \tabskip\@centering
    &{##}\tabskip\z@\cr}
\makeatother

\font\tenrmbf=cmbx12
\font\sevenrmbf=cmbx7
\font\fivermbf=cmbx5
\def\bfrm#1{{\textfont1=\tenrmbf\scriptfont1=\sevenrmbf
\scriptscriptfont1=\fivermbf
\mathchoice{\hbox{$\displaystyle#1$}}{\hbox{$\textstyle#1$}}
{\hbox{$\scriptstyle#1$}}{\hbox{$\scriptscriptstyle#1$}}}}
\def\P{\bfrm P}
\def\br{\be\begin{array}{|c|}\hline}
\def\er{\hline\end{array}\ee}

\begin{document}
\twocolumn[
\begin{flushright}{FIAN/TD-5/96\\
ITEP/TH-17/96}
\end{flushright}
{\Large\bf  WDVV Equations in Seiberg-Witten theory\\ and associative
algebras}\footnotemark{}
\par
\bigskip
A. Mironov\footnotemark{}\par
\medskip
Theory Department, P.N.Lebedev Physics
Institute, Moscow, Russia\\
and ITEP, Moscow, Russia\par
\smallskip\bigskip
\hspace{10pt}
This is a short review of the results on the associativity
algebras and WDVV equations found recently for the Seiberg-Witten solutions of
$N=2$ 4d SUSY gauge theories. The presentation is mostly based on the
integrable treatment of these solutions.\\
\bigskip]\footnotetext{Talk presented at the XXXIII Karpacz
Winter School "Duality -- String and Fields"}
\footnotetext{E-mail: mironov@lpi.ac.ru,
mironov@heron.itep.ru}

\paragraph{1. What is WDVV.} More than two years ago N.Seiberg and
E.Witten [1] proposed a new way
to deal with the low-energy effective actions of $N=2$ four-dimensional
supersymmetric gauge theories, both pure gauge theories (i.e. containing
only vector supermultiplet) and those with matter hypermultiplets. Among
other things, they have shown that the low-energy effective actions (the
end-points of the renormalization group flows) fit into universality classes
depending on the vacuum of the theory.  If the moduli space of these vacua is
a finite-\-dimensional variety, the effective actions can be essentially
described in terms of system with {\it finite}-dimensional phase space (\# of
degrees of freedom is equal to the rank of the gauge group), although the
original theory lives in a many-dimen\-si\-on\-al space-time.  These effective
theories turn out to be integrable. Integrable structure behind the
Seiberg-Witten (SW) appro\-ach has been found in [2] and later examined in
detail in [3].

The second important property of the SW fra\-me\-work
which merits the adjective
"topological" has been recently revealed in the series of papers [4,5]
and has
much to do with the associative algebras. Namely, it turns out that the
prepotential of SW theory satisfies a set of
Witten-Dijkgraaf-Verlinde-Verlinde (WDVV) equations. These equations has
been originally presented in [6] (in a different form, see below)
\be\label{wdvv}
F_iF_j^{-1}F_k=F_kF_j^{-1}F_i
\ee
where $F_i$'s are matrices with the matrix elements that are the third
derivatives of the unique function $F$ of many variables $a_i$'s
(prepotential in the SW theory) that given on a moduli space:
\be
\left(F_i\right)_{jk}\equiv {\partial^3 F\over \d a_i\d a_j\d a_k},
\ \ \ \ i,j,k=1,...,n
\ee
Although generally there is a lot of solutions to the matrix equations
(\ref{wdvv}), it is extremely non-trivial task to express all the matrix
elements through the only function $F$. In fact, there have been only known
the two different classes of the non-trivial solutions to the WDVV equations,
both being intimately related to the two-dimensional topological theories of
type A (quantum cohomologies [7]) and of type B ($N=2$ SUSY Landau-Ginzburg
(LG) theories that were investigated in a variety of papers, see, for
example, [8] and references therein). Thus, the existence of a new class of
solutions connected with the four-dimensional theories looks quite striking.
It is worth noting that both the two-dimensional topological theories and the
SW theories reveal the integrability structures related to the WDVV equations.
Namely, the function $F$ plays the role of the (quasiclassical)
$\tau$-function of some Whitham type hierarchy [8,2,3].

In this brief review, we will describe the results of papers [4,5] that deal
with the structure and origin of the WDVV equations in the SW theories.
To give some insight of the general
structure of the WDVV equations, let us consider
the simplest non-trivial examples of $n=3$ WDVV equations in topological
theories. The first example is the $N=2$ SUSY LG theory with the
superpotential $W'(\lambda)=\lambda^3-q$ [8]. In this case, the prepotential
reads as
\be
F=\2 a_1a_2^2+\2 a_1^2a_3+{q\over 2}a_2a_3^2
\ee
and the matrices $F_i$ (the third derivatives of the prepotential)
are\\[-5pt]
$$
F_1 = \left(\begin{array}{ccc}
0&0&1\\0&1&0\\1&0&0 \end{array}\right), \ \ \ \
F_2 = \left(\begin{array}{ccc}
0&1&0\\1&0&0\\0&0&q\end{array}\right),
$$
$$
F_3 = \left(\begin{array}{ccc}
1&0&0\\0&0&q\\0&q&0 \end{array}\right).
$$
One can easily check that these matrices do really satisfy the WDVV equations
(\ref{wdvv}).

The second example is the quantum cohomologies of C$\P ^2$. In this case, the
prepotential is given by the formula [7]\\[-3pt]
\be
F=\2 a_1a_2^2+\2 a_1^2a_3+\sum_{k=1}^{\infty}{N_ka_3^{3k-1}\over
(3k-1)!} e^{ka_2}
\ee
where the coefficients $N_k$ (describing the rational Gromov-Witten classes)
counts the number of the rational curves in
C$\P ^2$ and are to be calculated. Since the matrices $F$ have the form
$$
F_1 = \left(\begin{array}{ccc}
0&0&1\\0&1&0\\1&0&0\end{array}\right), \ \ \
F_2 = \left(\begin{array}{ccc}
0&1&0\\1&F_{222}&F_{223}\\0&F_{223}&F_{233}\end{array}\right),
$$
$$
F_3 = \left(\begin{array}{ccc}
1&0&0\\0&F_{223}&F_{233}\\0&F_{233}&F_{333}\end{array}\right)
$$
the WDVV equations are equivalent to the identity
\be
F_{333}=F_{223}^2-F_{222}F_{233}
\ee
which, in turn, results into the recurrent relation defining the coefficients
$N_k$:\\[-5pt]
$$
\frac{N_k}{(3k-4)!} = \sum_{a+b=k}
\frac{a^2b(3b-1)b(2a-b)}{(3a-1)!(3b-1)!}N_aN_b.
$$\\[-10pt]
The crucial feature of the presented examples is that, in both cases, there
exists a constant matrix $F_1$. Following [8], one can consider it as a flat
metric on the moduli space. In fact, in its original version, the WDVV
equations have been written in a slightly different form, that is, as
the associativity condition of some algebra. We will discuss this later, and
now just remark that, having distinguished (constant) metric $\eta\equiv F_1$,
one can naturally rewrite (\ref{wdvv}) as the equations\\[-7pt]
\be\label{Cas}
C_iC_j=C_jC_i
\ee\\[-13pt]
for the matrices $\left(C_i\right)_{jk}\equiv
\eta^{-1} F_i$, i.e. $C^j_{ik}=\eta^{jl}F_{ilk}$. Formula (\ref{Cas}) is
equivalent to (\ref{wdvv}) with $j=1$. Moreover, this particular relation is
already sufficient [4,5] to reproduce the whole set of the WDVV equations
(\ref{wdvv}).

Let us also note that, although the WDVV equations can be fulfilled only for
some specific choices of the coordinates $a_i$ on the moduli space, they
still admit any linear transformation. This defines the flat structures on
the moduli space, and we often call $a_i$ flat coordinates.

In fact, the existence of the flat metric is not necessary for
(\ref{wdvv}) to be true, how we explain below. Moreover, the SW theories give
exactly an example of such a case, where there is no distinguished constant
matrix. This matrix can be found in topological theories because of
existence their field theory interpretation where the unity operator is always
presented.

\paragraph{2. Perturbative SW prepotentials.}
Before going into the discussion of the WDVV equations for the complete SW
prepotentials, let us note that the leading perturbative part of them should
satisfy the equations (\ref{wdvv}) by itself (since the classical
quadratic piece does not contribute into the third derivatives). In each
case it can be checked by the straightforward calculation. On the other hand,
if the WDVV equations are fulfilled for perturbative prepotential, it is
a necessary condition for them to hold for complete prepotential.

To determine the one-loop perturbative prepotential from the field theory
calculation, let us note that
there are two origins of masses in ${\cal N}=2$ SUSY YM models:
first, they can be generated by vacuum values of the scalar $\phi$ from  the
gauge supermultiplet. For a supermultiplet in representation $R$
of the gauge group $G$ this contribution to the prepotential
is given by the analog of the Coleman-Weinberg formula
(from now on, we omit the classical part of the prepotential from all
expressions):
\be
F_R = \pm\frac{1}{4} \Tr_R \ \phi^2\log\phi,
\label{ColeWei}
\ee
and the sign is ``$+$'' for vector supermultiplets (normally they
are in the adjoint representation) and ``$-$'' for matter hypermultiplets.
Second, there are bare masses $m_R$ which should be added to $\phi$
in (\ref{ColeWei}). As a result, the general expression for the
perturbative prepotential is
$$
F = \frac{1}{4}\sum_{{vector}\atop{mplets}} \Tr_{A}
(\phi + M_nI_A)^2\log(\phi + M_nI_A) -
$$
$$
- \frac{1}{4}\sum_{{hyper}\atop{mplets}} \Tr_R
(\phi + m_RI_R)^2\log(\phi + m_RI_R) + f(m)
\label{PertuF}
$$
where the term $f(m)$ depending only on masses is not fixed by the
(perturbative) field theory but can be read off from the
non-perturbative description, and $I_R$ denotes the unit matrix in the
representation $R$.

As a concrete example, let us consider the $SU(n)$ gauge group. Then, say,
perturbative prepotential for the pure gauge theory acquires the
form\\[-10pt]
$$
F_V^{pert}={\f 4}\sum_{ij}
\left(a_i-a_j\right)^2\log\left(a_i-a_j\right)
$$
This formula establishes that when v.e.v.'s
of the scalar fields in the gauge supermultiplet are non-vanishing
(perturbatively $a_r$ are eigenvalues of the vacuum
expectation matrix  $\langle\phi\rangle$), the fields in the gauge multiplet
acquire masses $m_{rr'} = a_r - a_{r'}$ (the pair of indices $(r,r')$ label
a field in the adjoint representation of $G$). In the $SU(n)$ case,
the eigenvalues are subject to the condition $\sum_ia_i=0$.
Analogous formulas for the
adjoint matter contribution to the prepotential is
$$
F_A^{pert}=-{\f 4}\sum_{ij}
\left(a_i-a_j+M\right)^2\log\left(a_i-a_j+M\right)
$$
while the contribution of the fundamental matter reads as
$$
F_F^{pert}=-{\f 4}\sum_{i}
\left(a_i+m\right)^2\log\left(a_i+m\right)
$$

The perturbative prepotentials are discussed in detail in
[5]. In that paper is also contained the check of the WDVV equations for
these prepotentials. Here we just list the results.

{\bf i)} The WDVV equations always hold for the pure gauge theories
$F^{pert}=F_V^{pert}$.

{\bf ii)} If one considers the gauge supermultiplets interacting with the
matter hypermultiplets in the first fundamental representation with masses
$m_{\alpha}$ $F^{pert}=F_V^{pert}+rF_F^{pert}+Kf_F(m)$ (where $r$ and $K$
are some undetermined coefficients), the WDVV equations do not hold unless
$K=r^2/4$, $f_F(m)= {\f
4}\sum_{\alpha,\beta}\left(m_{\alpha}-m_{\beta}
\right)^2\log\left(m_{\alpha}-m_{\beta}\right)$, the masses being regarded
as moduli (i.e. the equations (\ref{wdvv}) contain the derivatives with
respect to masses).

{\bf iii)} If in the theory the adjoint matter hypermultiplets are presented,
i.e. $F^{pert}=F_V^{pert}+F_A^{pert}+f_A(m)$, the WDVV equations never hold.

From the investigation of the WDVV equations for the perturbative
prepotentials, one can learn the following lessons:
\begin{itemize}
\item
masses are to be regarded as moduli
\item
as an empiric rule, one may say that the WDVV
equations are satisfied by perturbative prepotentials which depend only on
the pairwise sums of the type $(a_i\pm b_j)$, where moduli $a_i$ and $b_j$ are
either periods or masses\footnote{This general rule can be
easily interpreted in D-brane terms, since the interaction of branes
is caused by strings between them. The pairwise structure $(a_i\pm
b_j)$ exactly reflects this fact, $a_i$ and $b_j$ should be identified with
the ends of string.}. This is the case for the models that contain either
massive matter hypermultiplets in
the first fundamental representation (or its dual), or massless
matter in the square product of those.
Troubles arise in all other situations because of the terms
with $a_i\pm b_j\pm c_k\pm\ldots$. (The inverse statement is wrong --
there are some exceptions when the WDVV equations hold despite the presence
of such terms -- e.g., for the exceptional groups.)
\item
at value $r=2$, like $a_i$'s lying in irrep of $G$, masses
$m_{\alpha}$'s can be regarded as lying in irrep of some $\widetilde G$ so
that if $G=A_n$, $C_n$, $D_n$, $\ \widetilde G=A_n$, $D_n$, $C_n$
accordingly.  This correspondence "explains" the form of the mass term in the
prepotential $f(m)$.
\end{itemize}
Our last example of the perturbative prepotential is related to the 5d SUSY
YM theory discussed by N.Nekrasov (see
the last reference of [3]). The theory is
considered compactified onto the circle of radius $R$, so that in
four-dimensions it can be seen as a gauge theory of the infinitely many
vector supermultiplets with masses $M_k=k/R$. For the sake of simplicity, we
put $R=1$. Then, the perturbative prepotentials in this theory reads as
\be\label{FARTC}
F^{pert}={1\over 4}\sum_{i,j}\left({1
\over 3}a_{ij}^3-{1\over 2}Li_3\left(e^{-2a_{ij}}\right)\right)-\\-
{n\over 4}\sum_{i>j>k}a_ia_ja_k
\ee\\[-10pt]
where $a_{ij}\equiv a_i-a_j$ and $Li_3(x)$ is the standard three-loga\-rithm.
The first sum in this expression tends to the usual logarithmic prepotential
$F_V^{pert}$ as $R\to 0$, while the second one vanishes. It deserves
mentioning that the second (cubic) term do not come from any field theory
calculation, but has a non-perturbative nature. It is similar to the
$U^3$-terms of the perturbative prepotential $F^{pert}$ of the heterotic
string [9]. The presence of these terms turns to be absolutely crucial for
the WDVV equations to hold in this case.

\paragraph{3. Associativity conditions.}
In the context of the two-dimensional LG topological theories, the
WDVV equations arose as associativity condition of some polynomial algebra.
We will prove below that the equations in the SW theories have the same
origin. Now we briefly remind the main ingredients of this approach in the
standard case of the LG theories.

In this case, one deals with the chiral ring formed by a set of polynomials
$\left\{\Phi_i(\lambda)\right\}$ and two co-prime (i.e. without common
zeroes) fixed polynomials $Q(\lambda)$ and $P(\lambda)$. The polynomials
$\Phi$ form the associative algebra with the structure constants $C_{ij}^k$
given with respect to the product defined by modulo $P'$:
\be
\Phi_i\Phi_j=C_{ij}^k\Phi_kQ'+(\ast)P'\longrightarrow C_{ij}^k\Phi_kQ'
\ee
the associativity condition being
\be
\left(\Phi_i\Phi_j\right)\Phi_k=\Phi_i\left(\Phi_j\Phi_k\right),
\ee
\be\label{ass}
\hbox{ i.e. }\ \ \
C_iC_j=C_jC_i,\ \ \ \left(C_i\right)_k^j=C_{ik}^j
\ee
Now, in order to get from these conditions the WDVV equations, one needs to
choose properly the flat moduli [8]:\\[-5pt]
$$
a_i=-{n\over i(n-i)}\hbox{res}\left(P^{i/n}dQ\right),\ \ \ n=\hbox{ord} (P)
$$
Then, there exists the prepotential whose third derivatives are given by
the residue formula
\be\label{res}
F_{ijk}=
\stackreb{P' = 0}{\hbox{res}}
\frac{\Phi_i\Phi_j\Phi_k}{P'}
\ee
On the other hand, from the associativity condition (\ref{ass}) and residue
formula (\ref{res}), one obtains that\\[-10pt]
\be\label{im}
F_{ijk}=\left(C_i\right)_j^lF_{Q'lk},\ \ \hbox{ i.e. }\ \ C_i=F_iF^{-1}_{Q'}
\ee
Substituting this formula for $C_i$ into (\ref{ass}), one finally reaches the
equations of the WDVV type. The choice $Q'=\Phi_l$ gives the standard
equations (\ref{wdvv}). In two-dimensional topological theories, there is
always the unity operator that corresponds to $Q'=1$ and leads to the
constant metric $F_{Q'}$.

Thus, from this short study of the WDVV equations in the LG theories, we can
get three main ingredients necessary for these equations to hold. These are:
\begin{itemize}
\item
associative algebra
\item
flat moduli (coordinates)
\item
residue formula
\end{itemize}
We will show that in the SW theory only the first ingredient requires
a non-trivial check.

\paragraph{4. SW theories and integrable systems.}
Now we turn to the WDVV equations in the SW construction [1] and show how
they are related to integrable system underlying the corresponding SW theory.
The most
important result of [1], from this point of view, is that the moduli space of
vacua and low energy effective action in SYM theories
are completely given by the following input data:
\begin{itemize}
\item
Riemann surface ${\cal C}$
\item
moduli space ${\cal M}$ (of the curves ${\cal C}$)
\item
meromorphic 1-form $dS$ on ${\cal C}$
\end{itemize}
How it was pointed out in [2,3], this input can be naturally described
in the framework of some underlying integrable system. Let us consider a
concrete example -- the $SU(n)$ pure gauge SYM theory that
can be described by the periodic Toda chain with $n$ sites.
This integrable system is
entirely given by the Lax operator
\be\label{Lax}
L(w)=\left(\begin{array}{cccc}
p_1 & e^{q_1-q_2} & & w\\
e^{q_1-q_2} & p_2 & \vdots&\\
&\ldots&\ddots &\vdots\\
{\f w}&& \ldots&p_n
\end{array}\right)
\ee
The Riemann surface ${\cal C}$ of the SW data is nothing but the spectral
curve of the integrable system, which is given by the equation\\[-7pt]
$$
\det\left(L(w)-\lambda\right)=0
$$
Taking into account (\ref{Lax}), one can get from this formula the
equation\\[-7pt]
\begin{equation}\label{scurve}
w+{\f w}=P\left(\lambda\right)=\prod_{i=1}^n\left(\lambda-\lambda_i\right),
\ \ \ \sum_i\lambda_i=0
\end{equation}
where the ramification points $\lambda_i$ are Hamiltonians (integrals
of motion) parametrizing the moduli space ${\cal M}$ of the spectral curves.
The replace $Y\equiv w-1/w$ transforms the curve (\ref{scurve}) to the
standard hyperelliptic form $\ Y^2=P^2-4$, the genus of the curve
being $n-1$.                                 

As to the meromorphic 1-form $dS=\lambda{dw\over
w}=\lambda {dP\over Y}$, it is just the shorten action "$pdq$"
along the non-contractible contours on the Hamiltonian
tori. Its defining property is that the derivatives of
$dS$ with respect to the moduli (ramification points)
are holomorphic differentials on the spectral curve.
                      
Now let us describe the general integrable fra\-me\-work for the
SW construction and start with the theories without matter
hypermultiplets. First,                                                                                                                                                                                                                                                                                                                                                                                                                                                                      matter hypermultiplets. First,
we introduce bare spectral curve $E$ that is torus
$y^2=x^3+g_2x^2+g_3$ for the UV finite
SYM theories with the associated holomorphic 1-form
$d\omega=dx/y$. This bare spectral curve degenerates into the
double-punctured sphere (annulus) for the asymptotically free theories: $x\to
w+1/w$, $y\to w-1/w$, $d\omega=dw/w$. On this bare curve, there is given a
matrix-valued Lax operator $L(x,y)$. The dressed spectral curve is
defined from the formula $\det(L-\lambda)=0$. This spectral curve is a
ramified covering of $E$ given by the equation
\be
{\cal P}(\lambda;x,y)=0
\ee
In the case of the gauge group  $G=SU(n)$, the function ${\cal P}$ is a
polynomial of degree $n$ in $\lambda$.

Thus, the moduli space ${\cal M}$ of the spectral curve is given just  by
coefficients of ${\cal P}$.
The generating 1-form $dS \cong \lambda d\omega$ is meromorphic on
${\cal C}$ (hereafter the equality modulo total derivatives
is denoted by ``$\cong$'').

The prepotential and other "physical" quantities are defined in terms of the
cohomology class of $dS$:\\[-10pt]
\be
a_i = \oint_{A_i} dS,\ \ \ \ a_i^D\equiv {\d F\over\d a_i}=\oint_{B_i}dS,\\
\ A_I \circ B_J = \delta_{IJ}.
\label{defprep}
\ee
The first identity defines here the appropriate flat moduli, while the second
one -- the prepotential. The derivatives of the generating differential
$dS$ give holomorphic 1-differentials:
\be
{\d dS\over \d a_i}=d\omega_i
\ee
and, therefore, the second derivative of the prepotential is the period
matrix of the curve ${\cal C}$:
\be
{\d^2F\over\d a_i\d a_j}=T_{ij}
\ee
The latter formula allows one to identify prepotential with logarithm of the
$\tau$-function of Whitham hierarchy [8]: $F=\log\tau$.

So far we reckoned without massive hypermultiplets.
In order to include them, one just needs to consider the
surface ${\cal C}$ with punctures. Then, the masses are proportional to
residues of $dS$ at the punctures, and the moduli space has to be extended to
include these mass moduli. All other formulas remain in essence the same
(see [4,5] for more details).

By the present moment, the correspondence between SYM theories and integrable
systems is built through the SW construction in most of known cases that
are collected in the table\footnote{In the table we considered only the
classical groups (see below).}.

Note that the only theory, when the SW approach is
applied but the corresponding integrable system still remains unknown is the
UV finite SYM theory with the matter hypermultiplets in the first fundamental
representation.

\newpage
\twocolumn[
{\large {\bf Table.}
SUSY gauge theories $\Longleftrightarrow$ integrable systems
correspondence}\par\medskip
\begin{center} \begin{tabular}{|c|c|c|c|c|} \hline SUSY     &
4d pure gauge   & 4d SYM with & 4d SYM with & 5d pure gauge\\ physical &  SYM
theory,    & fundamental & adjoint matter & SYM theory\\ theory   & gauge
group $G$ & matter      &       &\\ \hline Underlying & Toda chain  &
Rational  & Calogero-Moser & Relativistic\\ integrable & for the dual  & spin
chain &  system & Toda chain\\ system     & affine ${\hat G}^{\vee}$ & of XXX
type & & \\ \hline Bare     &        &        &       & \\ spectral & sphere
& sphere & torus & sphere \\ curve    &        &        &       & \\ \hline
Dressed  &   &  &  & \\
spectral & hyperelliptic & hyperelliptic & non-hyperelliptic &
hyperelliptic \\
curve &        &        &       & \\
\hline
Generating &&&&\\
meromorphic & $\lambda{dw\over w}$ & $\lambda{dw\over w}$ &
$\lambda{dx\over y}$ &  $\log\lambda{dw\over w}$ \\
1-form $dS$&&&&\\
\hline \end{tabular}
\end{center}{}
\vspace{10pt}
]
\noindent
To complete this table, we describe the dressed spectral curves in each case
in more explicit terms. Let us note that in all but adjoint matter cases the
curves are hyperelliptic and can be described by the general formula
\be
{\cal P}(\lambda, w) = 2P(\lambda) - w - \frac{{\cal Q}(\lambda)}{w}
\label{curven}
\ee
Here $P(\lambda)$ is characteristic polynomial of the
algebra $G$ itself, i.e.
\be
P(\lambda) = \det(G - \lambda I) =
\prod_i (\lambda - \lambda_i)
\ee
where determinant is taken in the first fundamental representation
and $\lambda_i$'s are the eigenvalues of the algebraic element $G$.
For the pure gauge theories with the classical groups,
${\cal Q}(\lambda)=\lambda^{2s}$ and
\be
A_{n-1}:\ \ \ P(\lambda) = \prod_{i=1}^{n}(\lambda - \lambda_i), \ \ \
\ s=0;\\
B_n:\ \ \ P(\lambda) = \lambda\prod_{i=1}^n(\lambda^2 - \lambda_i^2),
\ \ \ s=2;\\
C_n:\ \ \ P(\lambda) = \prod_{i=1}^n(\lambda^2 - \lambda_i^2),
\ \ \ s=-2;\\
D_n:\ \ \ P(\lambda) = \prod_{i=1}^n(\lambda^2 - \lambda_i^2),
\ \ \ s=2
\label{charpo}
\ee
For exceptional groups, the curves arising from the characteristic
polynomials of the dual affine algebras do not acquire the hyperelliptic
form. Therefore, in this case, the line "dressed spectral curve" in the table
has to be corrected.

In order to include $n_F$ massive hypermultiplets in the
first fundamental representation one can just change
$\lambda^{2s}$ for ${\cal Q}(\lambda) = \lambda^{2s}
\prod_{\iota = 1}^{n_F} (\lambda - m_\iota)$ if $G=A_n$
and for ${\cal Q}(\lambda) = \lambda^{2s} \prod_{\iota =
1}^{n_F}(\lambda^2 - m^2_\iota)$ if $G=B_n,C_n,D_n$ [10].

At last, the 5d theory is just described by ${\cal
Q}(\lambda)$ $=\lambda^{n/2}$.

In the Calogero-Moser case, the spectral curve is non-hyperelliptic, since the
bare curve is elliptic. Therefore, it can be described as some covering of
the hyperelliptic curve. We do not go into
further details here, just referring to [5].

\paragraph{5. WDVV equations in SW theories.}
As we already discussed, in order to derive the WDVV equations along the line
used in the context of the LG theories, we need three crucial ingredients:
flat moduli, residue formula and associative algebra. However, the first two
of these are always contained in the SW construction provided the
underlying integrable system is known. Indeed, one can derive (see [4,5]) the
following residue formula
\be\label{resSW}
F_{ijk}=
\stackreb{d\omega= 0}{\hbox{res}}
\frac{d\omega_id\omega_jd\omega_k}{d\omega d\lambda}
\ee
where the proper flat moduli $a_i$'s are given by formula (\ref{defprep}).
Thus, the only point is to be checked is the existence of the associative
algebra. The residue formula (\ref{resSW}) hints that this algebra is to be
the algebra $\Omega^1$ of the holomorphic differentials $d\omega_i$. In the
forthcoming discussion we restrict ourselves to the case of pure gauge
theory, the general case being treated in complete analogy.

Let us consider the algebra $\Omega^1$ and fix three differentials $dQ$,
$d\omega$, $d\lambda\ \in\Omega^1$. The product in this algebra is given
by the expansion
\be\label{prod}
d\omega_id\omega_j=C^k_{ij}d\omega_kdQ+(\ast)d\omega+(\ast)d\lambda
\ee
that should be factorized over the ideal spanned by the differentials
$d\omega$ and $d\lambda$.
This product belongs to the space of quadratic holomorphic differentials:
\be
\Omega^1\cdot\Omega^1\in\Omega^2\cong\Omega^1\cdot\left(dQ
\oplus d\omega\oplus d\lambda\right)
\ee
Since the dimension of the space of quadratic holomorphic differentials is
equal to $3g-3$, the l.h.s. of (\ref{prod}) with arbitrary $d\omega_i$'s is
the vector space of dimension $3g-3$. At the same time, at the
r.h.s. of (\ref{prod}) there are $g$ arbitrary coefficients $C_{ij}^k$ in the
first term (since there are exactly so many holomorphic 1-differentials that
span the arbitrary holomorphic 1-differential $C_{ij}^kd\omega_k$), $g-1$
arbitrary holomorphic differentials in the second term (one differential
should be subtracted to avoid the double counting) and $g-2$ holomorphic
1-differentials in the third one. Thus, totally we get that the r.h.s. of
(\ref{prod}) is spanned also by the basis of dimension $g+(g-1)+(g-2)=3g-3$.

This means that the algebra exists in the general case of the SW construction.
However, this algebra generally is not associative. This is because, unlike
the LG case, when it was the
algebra of polynomials and, therefore, the product
of the two belonged to the same space (of polynomials), product in the algebra
of holomorphic 1-differentials no longer belongs to the same space but to the
space of quadratic holomorphic differentials. Indeed, to check associativity,
one needs to consider the triple product of $\Omega^1$:
\be\label{assSW}
\Omega^1\cdot\Omega^1\cdot\Omega^1\in\Omega^3=
\Omega^1\!\cdot
\left(dQ\right)^2\oplus\Omega^2\!\cdot d\omega\oplus\Omega^2
\!\cdot d\lambda
\ee
Now let us repeat our calculation: the dimension of the l.h.s. of this
expression is $5g-5$ that is the dimension of the space of holomorphic
3-differentials. The dimension of the first space in expansion of the r.h.s.
is $g$, the second one is $3g-4$ and the third one is $2g-4$. Since
$g+(3g-4)+(2g-4)=6g-8$ is greater than $5g-5$ (unless $g\le 3$), formula
(\ref{assSW}) {\bf does not} define the unique expansion of the triple
product of $\Omega^1$ and, therefore, the associativity spoils.

The situation can be improved if one considers the curves with additional
involutions. As an example, let us consider the family of hyperelliptic
curves: $y^2=Pol_{2g+2}(\lambda)$. In this case, there is the involution,
$\sigma:\ y\to -y$ and $\Omega^1$ is spanned by the $\sigma$-odd holomorphic
1-differentials ${x^{i-1}dx\over y}$, $i=1,...,g$. Let us also note that both
$dQ$ and $d\omega$ are $\sigma$-odd, while $d\lambda$ is $\sigma$-even. This
latter fact means that $d\lambda$ can be only meromorphic unless there are
punctures on the surface (which is, indeed, the case in the presence of the
mass hypermultiplets). Thus, formula (\ref{prod}) can be replaced by that
without $d\lambda$
\be\label{prodhe}
\Omega^2_+=\Omega^1_-\cdot dQ\oplus\Omega^1_-\cdot d\omega
\ee
where we expanded the space of holomorphic 2-differentials into the parts
with definite $\sigma$-parity: $\Omega^2=\Omega^2_+\oplus\Omega^2_-$, which
are manifestly given by the differentials ${x^{i-1}(dx)^2\over y^2}$,
$i=1,...,2g-1$ and ${x^{i-1}(dx)^2\over y}$, $i=1,...,g-2$ respectively.
Now it is easy to understand that the dimensions of the l.h.s. and r.h.s. of
(\ref{prodhe}) coincide and are equal to $2g-1$.

Analogously, in this case, one can check the associativity. It is given by the
expansion
\be
\Omega^3_-=\Omega_-^1\cdot \left(dQ\right)^2\oplus\Omega_+^2\cdot d\omega
\ee
where both the l.h.s. and r.h.s. have the same dimension: $3g-2=g+(2g-2)$.
Thus, the algebra of holomorphic 1-differentials on hyperelliptic curve
is really associative. This completes the proof of the WDVV equations in this
case.

Now let us briefly list when there exist the associative algebras basing
on the spectral curves discussed in the previous section. First of all, it
exists in the theories with the gauge group $A_n$, both in the pure gauge
4d and 5d theories and in the theories with fundamental matter, since, in
accordance with s.4, the corresponding spectral curves are hyperelliptic ones
of genus $n$.

Equally the theories with the gauge groups $SO(n)$ or $Sp(n)$ are described
by the hyperelliptic curves. The curves, however, are of higher genus
$2n-1$. This would naively destroy all the reasoning of this section.
The arguments,
however, can be restored by noting that the corresponding curves (see
(\ref{charpo})) have yet {\bf another} involution, $\rho:\
\lambda\to-\lambda$. This allows one to expand further the space of
holomorphic differentials into the pieces with definite $\rho$-parity:
$\Omega^1_-=\Omega^1_{--}\oplus\Omega^1_{-+}$ etc. so that the proper algebra
is generated by the differentials from $\Omega^1_{--}$. One can easily check
that it leads again to the associative algebra.

Consideration is even more tricky for the exceptional groups, when the
corresponding curves are non-hyperelliptic. However, additional involutions
allow one to get associative algebras in these cases too.

The situation is completely different in the adjoint matter case that is
described by the Calogero-Moser integrable system. Since, in this case, the
curve is non-hyperelliptic and has no evident involutions, one needs to
include into consideration both the differentials $d\omega$ and $d\lambda$
for algebra to exist. However, under these circumstances, the algebra is no
longer associative how it was demonstrated above. This can be done also by
direct calculation for several first values of $n$ (see [5]).

\paragraph{6. Concluding remarks.}
To conclude these short notes, let us emphasize that, along with already
mentioned problem of lack of the WDVV equations for the Calogero-Moser
system, no counterpart of the WDVV equations is also known yet for the
Calabi-Yau manifolds (the naive one is just empty [7]). The way to resolve
these problems might be to construct higher associativity conditions like it
has been done by E.Getzler in the elliptic case [11], that is to say, for the
elliptic Gromov-Witten clas\-ses. It may deserve mentioning that the WDVV
equations in the type A topological theories themselves do still wait for the
explanation in terms of associative algebras. All these problems are to be
resolved in order to establish to what extent there is a really deep reason
for the WDVV equations to emerge.

I am grateful to E.Akhmedov, A.Gorsky, S.Gu\-kov, I.Krichever,
A.Losev, A.Marshakov, A.Moro\-zov, N.Nekrasov and I.Polyubin
for useful discussions. This work was supported by grant
INTAS-RFBR-95-0690.

\paragraph{\large {\bf References}}
\begin{enumerate}
\item
N.Seiberg, E.Witten, Nucl.Phys. {\bf B426} (1994) 19; 484
\item
A.Gorsky et al,
Phys.Lett. {\bf B355} (1995) 466
\item
E.Martinec, N.Warner, hepth/9509161\\
T.Nakatsu, K.Takasaki, hepth/9509162\\
A.Marshakov, Mod.Phys.Lett. {\bf A11} (1996) 1169\\
C.Ann, S.Nam, hepth/9603028\\
A.Gorsky et al., Phys.Lett. {\bf B380} (1996)
75\\
A.Gorsky et al., hepth/9604078\\
R.Donagi, E.Witten, hepth/9510101\\
E.Martinec, hepth/9510204\\
A.Gorsky, A.Marshakov, Phys.Lett. {\bf B375} (1996) 127\\
E.Martinec, N.Warner, hepth/9511052\\
H.Itoyama, A.Morozov, hep-th/9601168;\\ 9511126; 9512161\\
I.Krichever, D.Phong, hepth/9604199\\
N.Nekrasov, hepth/9609219
\item
A.Marshakov et al,
hepth/9607109\\
A.Marshakov et al, hepth/9701014
\item
A.Marshakov et al, hepth/9701123
\item
E.Witten, Surv.Diff.Geom. {\bf 1} (1991) 243\\
R.Dijkgraaf, E.Verlinde, H.Verlinde,\\ Nucl.Phys.
{\bf B352} (1991) 59
\item
Yu.Manin, {\sl Frobenius manifolds, quantum cohomology and moduli spaces},
MPI, 1996\\
M.Kontsevich, Yu.Manin, Comm.Math.Phys. {\bf 164} (1994) 525
\item
I.Krichever, Comm.Pure Appl.Math. \\{\bf 47} (1994) 437\\
B.Dubrovin,  Nucl.Phys. {\bf B379} (1992) 627
\item
J.Harvey, G.Moore, hepth/9510182
\item
A.Hanany, Y.Oz, hepth/9505075\\
P.Argyres, A.Shapere, hepth/9509175\\
A.Hanany, hepth/9509176\\
A.Brandhuber, K.Landsteiner, hepth/9507008\\
U.H.Danielsson, B.Sundborg, hepth/9504102
\item
L.Caporaso, J.Harris, alg-geom/9608025\\
E.Getzler, alg-geom/9612004
\end{enumerate}

\end{document}